\documentclass[
 aps,
 pra,
 reprint,
 superscriptaddress,
 amsmath,
 amssymb,
 floatfix
]{revtex4-2}

\usepackage[utf8]{inputenc}
\usepackage[T1]{fontenc}
\usepackage{graphicx}
\usepackage{xcolor}
\usepackage{bm}
\usepackage{braket}
\usepackage{dcolumn}
\usepackage{chemformula}
\usepackage{physics}

\def\beit{\begin{itemize}}
\def\eit{\end{itemize}}

\begin{document}

\title{Polyatomic Thermal Radiative Dissociation in Microcavities}

\author{Enes Suyabatmaz}
\affiliation{Department of Physics, Emory University, Atlanta, Georgia 30322, USA}

\author{Tuan Nguyen}
\affiliation{Department of Chemistry, Emory University, Atlanta, Georgia 30322, USA}

\author{Raphael F. Ribeiro}
\email{raphael.ribeiro@emory.edu}
\affiliation{Department of Chemistry, Emory University, Atlanta, Georgia 30322, USA}

\date{\today}

\begin{abstract}
Thermal infrared radiative dissociation activates molecules via successive absorption of ambient thermal photons until the internal energy reaches a dissociation threshold. Because these radiative
transition rates depend on the electromagnetic density of states, structured infrared environments provide a route to control thermal unimolecular dissociation. Here we develop a state-resolved
master-equation framework for polyatomic BIRD in a planar Au/MgO multilayer cavity, where the reactive cluster \(\mathrm{(H_2O)_2Cl^-}\) is studied. The resonator changes the dissociation kinetics through the density of states sampled by anharmonic fundamental, overtone, and combination
transitions. We show
that MgO surface phonon polaritons produce strong
near-field enhancements in the central vacuum reaction region of a microcavity. We find that short
microcavities with thick polar crystal layers yield the largest thermal dissociation enhancements due
to evanescent surface phonon polariton contributions. We further include
collisions with a methane bath gas and show that microcavity DOS engineering
shifts the crossover between radiative and
collisional activation. These results establish
Reststrahlen-band  engineering as a strategy for controlling polyatomic thermal radiative dissociation in microcavities.
\end{abstract}

\maketitle

\section{Introduction}

%Controlling chemical reactivity with confined light has %opened a
%complementary axis to traditional thermodynamic and chemical levers such
%as temperature, pressure, and catalyst composition. 

Infrared (IR) microcavities reshape the electromagnetic environment experienced by
molecules and can modify radiative transition rates by changing the
available photonic density of states (DOS) \cite{kavokin2017microcavities,
e.m.purcell1946,barnes_classical_2020}. In many experiments, this idea
has been explored in vibrational strong coupling, where molecular
vibrations hybridize with confined electromagnetic modes to form vibrational
polaritons \cite{vinogradov1992,shalabney_coherent_2015,
simpkins_spanning_2015}. A growing body of theory and experiment has
suggested that the vibrational strong coupling regime in infrared microcavities can influence reaction kinetics,
although the microscopic origin, limits, and generality of these effects remain under debate \cite{thomas2016,thomas2019,hirai2020,ahn2023,
galego2019cavity,campos2019resonant,zhdanov2020vacuum,
wang2022chemical,lindoy2022resonant,schafer2022shining,
sun2022suppression,Vega2025,lai2024non,
moiseyev2024conditionsenhancementchemicalreactions,simpkins2021mode,campos2023swinging,
MontilloVega_Ying_Huo_2024,sanchez2022theoretical,
Lai2026AnalyticRateTheoryPolaritons}.

Gas-phase blackbody infrared radiative dissociation (BIRD) provides a
direct platform for isolating DOS-mediated microcavity effects. BIRD was
anticipated in early radiation-driven unimolecular theories
\cite{perrin1970atomes,lindemann1922discussion,daniels1928radiation}
and later established experimentally as a dominant low-pressure
activation mechanism for trapped molecular ions\cite{king1984chemical,
thoelmann1994spontaneous,dunbar1995zero,schnier1996blackbody,
dunbar1998activation,dunbar2004birdreview}. In BIRD, a molecule or ion
absorbs thermal IR photons through a sequence of vibrational transitions until its internal energy reaches the
dissociation threshold. Because the absorption and emission rates are
set by Einstein coefficients and by the electromagnetic DOS at each
transition frequency, replacing free space with a structured IR
environment provides a direct route to modify the radiative transition network. We recently provided a demonstration of this type of DOS-mediated modulation of BIRD rates for
a diatomic molecule in a weakly coupled microcavity and in a polaritonic environment \cite{suyabatmaz2025polaritonic}.
The present work expands this idea to a polyatomic ion, where the activation pathway is distributed over a dense anharmonic vibrational manifold.

For polyatomic molecules, BIRD is intrinsically high-dimensional. Fundamental transitions, overtones, and combination bands form coupled vibrational ladders, and mechanical and electrical anharmonicities shift both the transition frequencies and the transition
strengths\cite{gilbert_smith_1990,klippenstein2022spiers,
dunbar1989ir_cooling,dunbar1992ir_cooling_review,
beyer_swinehart1973,stein_rabinovitch1973}. Intramolecular vibrational energy redistribution further couples mode-specific excitation to the total internal energy flow toward dissociation \cite{leitner2006a,
leitner2018molecules}. As a result, microcavity control of polyatomic BIRD cannot generally be reduced to tuning one resonance to one molecular line. Instead, the rate is controlled by how the structured electromagnetic DOS overlaps with the subset of anharmonic transitions
that contribute significantly to the reactive flux.
Properly addressing these kinetically important transitions requires a state-resolved treatment of both the molecular vibrational manifold and
the microcavity-modified radiative rates\cite{barker2001multiwell,
glowacki2012mesmer}.

Here we focus on a treatment of polyatomic BIRD in a planar multilayer microcavity formed by metallic mirrors and polar crystal slabs Fig.~\ref{fig:multilayer_schematic}. The reactive chemical species remains weakly coupled to the electromagnetic field throughout, i.e., the molecular quantum states are not hybridized with microcavity
modes. The infrared resonator modifies the kinetics only by changing the local electromagnetic DOS sampled by the molecular transitions. The key physical ingredient is the Reststrahlen response of the polar crystal layer. Between the transverse and longitudinal optical phonon frequencies of a polar crystal, the real part of the dielectric function becomes negative,
allowing surface phonon polaritons to form at polar
interfaces \cite{raether1988surface,joulain2005surface,gubbin2022surface,chance1978molecular,agarwal1975qed_i,agarwal1975qed_iv,
wylie1984qed,wylie1985qed,barnes1998fluorescence,novotny_hecht_2012,
barnes_classical_2020}
These surface modes are evanescent in the direction normal to the interface and can strongly enhance the near-field electromagnetic DOS \cite{fuchs1965ionic_slab,kliewer1966slab_nonradiative,
kliewer1966slab_radiative,economou1969thin_films}. %For finite polar slabs and multilayers, the surface modes on different
%interfaces hybridize, producing geometry-dependent phonon-polariton branches whose frequencies and linewidths depend on the polar-layer thickness, spacer thickness, cavity length, and material losses

We develop a coarse-grained Pauli master-equation framework for
\(\mathrm{(H_2O)_2Cl^-}\) that combines anharmonic vibrational state
counting, state-resolved absorption and emission rates, and
microcavity-modified electromagnetic DOS factors \cite{pauli1928festschrift,
vankampen2007,gardiner2009,gilbert_smith_1990,barker2001multiwell,
glowacki2012mesmer,dunbar1995zero,dunbar2004birdreview}. The molecular data includes fundamental, overtone, and combination-band transitions,
so the model resolves how different classes of anharmonic transitions respond to the structured infrared environment. We then apply this framework to a family of Au/MgO multilayer geometries in which both the
microcavity length and the MgO layer thickness are varied. This allows us to determine how Reststrahlen-band surface phonon polaritons, evanescent near fields, and guided modes jointly control the BIRD rate.

We include collisions with a methane bath gas to determine when the microcavity-induced changes to radiative activation are observable in the presence of
collisional energy transfer. At very low pressures, the BIRD rate is governed by blackbody radiation, whereas at higher pressures collisional activation dominates and the overall kinetics become less sensitive to the electromagnetic DOS. This competition
defines a microcavity-modified radiation-collision crossover, providing a practical criterion for identifying radiation-dominated, mixed, and collision-dominated regimes in polyatomic BIRD kinetics.

This article is organized as follows. Section~II introduces the zero pressure master equation model, the Green-tensor computation of the microcavity local density of states, and the treatment of collisional energy transfer.
Section~III presents \(\mathrm{(H_2O)_2Cl^-}\) dissociation rates and microcavity length and MgO thickness
dependence of rate enhancements, the corresponding
phonon-polariton DOS and dispersion analysis, and the analysis of pressure dependent crossover from
radiative to collisional activation. Section~IV summarizes our main conclusions on how Reststrahlen-band DOS engineering can be used to control molecular thermal infrared radiative dissociation.

\section{Computational Methods}
\subsection{Pauli Master equation approach}

We model zero-pressure blackbody infrared radiative dissociation using a coarse-grained Pauli master equation in the internal
vibrational energy space \cite{dunbar1989ir_cooling,dunbar1995zero,
dunbar2004bird_review,price1997aminoacid,salzburger2024mem,
salzburger2022multiplewell, suyabatmaz2025polaritonic}. The molecular population is propagated on
an energy grid, while thermal radiation absorption, stimulated emission, spontaneous emission,
and irreversible dissociation are included as
transitions between energy grains. The total vibrational energy is
binned with a uniform grain size $\Delta E$, with bin centers $
E_i=\left(i+\frac{1}{2}\right)\Delta E .
\label{eq:bin-centers}$
Previous BIRD studies found that
$\Delta E=100~\mathrm{cm^{-1}}$ is sufficient for the overall
dissociation rate \cite{dunbar1995zero,price1997aminoacid}. Here we use a finer graining of $\Delta E=20~\mathrm{cm^{-1}}$ in order to resolve
lower frequency anharmonic hot band steps and reduce binning artifacts when assigning state-dependent transitions to final energy grains.

We collect the coarse-grained populations into a column vector
$\mathbf p(t)=(p_0(t),p_1(t),\ldots)^{\mathsf T}$, where $p_i(t)$ is
the population in energy bin $i = 0,1,2,...,i_{\text{max}}$. The total rate for population transfer
from bin $i$ to bin $j$ is denoted by $k_{i\to j}$. The population
dynamics are
\begin{equation}
\frac{\mathrm{d} p_i}{\mathrm{d}t}
=
\sum_{j\neq i} k_{j\to i}p_j
-
p_i\sum_{j\neq i} k_{i\to j}
-
k_{\rm diss}(E_i)p_i .
\label{eq:population-bin}
\end{equation}
The first term describes population gained from other energy bins, the
second describes radiative loss from bin $i$, and the last term is
the irreversible dissociation sink. Equation~\eqref{eq:population-bin}
can be written as
\begin{equation}
\frac{d\mathbf p(t)}{dt}
=
-\mathbf J\,\mathbf p(t),
\label{eq:j-master}
\end{equation}
where $\mathbf J$ is the coarse-grained transition matrix.
%In the absence of dissociation, the radiative part of the transition matrix conserves total population, which provides a numerical check on
%the matrix construction.

We selected the chloride-water dimer as a benchmark system
$(\mathrm{H_2O})_2\mathrm{Cl}^-$, because its zero pressure thermal radiation induced dissociation was measured and modeled by Dunbar et al \cite{dunbar1995zero}. The dissociation channel is
\begin{equation}
(\mathrm{H_2O})_2\mathrm{Cl}^-
\rightarrow
(\mathrm{H_2O})\mathrm{Cl}^-+\mathrm{H_2O}.
\label{eq:h2o2cl-dissociation}
\end{equation}
The dissociation threshold $E_0$ defines the absorbing boundary of the vibrational manifold. Dunbar et al. showed that RRKM theory gives very fast dissociation at the first grain above $E_0$ for these small cluster ions, larger than
$10^7~\mathrm{s^{-1}}$ \cite{dunbar1995zero, dunbar2004bird_review}. Therefore the actual
above-threshold dissociation rate values are irrelevant as long as they are very large (over $10^7~\mathrm{s^{-1}}$). Equivalently, the transition matrix may be truncated at the
threshold energy. We use this absorbing-boundary formulation by truncating the transition matrix at $E_0$.

With the absorbing boundary included, the long time decay of the bound population is exponential and the dissociation rate constant is obtained from the smallest
positive eigenvalue of the transition  matrix \cite{valance1966theoretical, suyabatmaz2025polaritonic}
\begin{equation}
k_{\rm BIRD}
=
\lambda_{\min}(\mathbf J).
\label{eq:rate-from-j}
\end{equation}

\subsection{Molecular model}
We construct a model for the vibrational manifold of $(\mathrm{H_2O})_2\mathrm{Cl}^-$ from a PBE0-D4/def2-TZVPD level second order vibrational perturbation theory (VPT2) computation performed with
ORCA \cite{adamo1999pbe0,caldeweyher2019d4,weigend2005def2,
rappoport2010def2diffuse,neese2020orca,neese2022orca,
barone2014anharmonic,franke2021vpt2}. %The VPT2 computation used the RIJCOSX 
%approximation with the def2/J auxiliary basis, DEFGRID3, and VeryTightSCF
%convergence \rev{[...]}. 
These choices ensure an appropriate anharmonic vibrational manifold and IR line strengths allowing us to incorporate anharmonic frequency shifts,
mode coupling, overtones, and combination bands into our kinetic model (see SI for additional information).

For a vibrational state
$\mathbf n=(n_1,n_2,\ldots,n_{15})$, 
the coupled rigid-rotor
anharmonic-oscillator energy can be written as\cite{barone2014anharmonic, franke2021vpt2, chen2026rrao_anharmonic}
\begin{equation}
\begin{split}
G(\mathbf n)
&=
G_0+
\sum_{a=1}^{15}
\omega_a\left(n_a+\frac{1}{2}\right) \\
&\quad+
\sum_{a=1}^{15}\sum_{b=a}^{15}
\chi_{ab}
\left(n_a+\frac{1}{2}\right)
\left(n_b+\frac{1}{2}\right),
\end{split}
\label{eq:vpt2-energy-full}
\end{equation}
where $\omega_a$ are the harmonic frequencies and $\chi_{ab}$ are real anharmonic constants. The relative excitation energy adopted in the master equation formalism is measured from the zero-point state,
\begin{equation}
E(\mathbf n)=G(\mathbf n)-G(\mathbf 0).
\label{eq:vpt2-relative-energy}
\end{equation}
Equivalently\cite{mendolicchio2023curvilinear},
\begin{equation}
\begin{split}
E(\mathbf n)
&=
\sum_{a=1}^{15}\omega_a n_a
+
\sum_{a=1}^{15}\chi_{aa}(n_a^2+n_a) \\
&\quad+
\sum_{a<b}^{15}
\chi_{ab}
\left[
n_a n_b+\frac{n_a+n_b}{2}
\right].
\end{split}
\label{eq:vpt2-relative-energy-expanded}
\end{equation}
This equation shows the excitation frequency of a mode depends on the excitation state of the others via the
off-diagonal anharmonic constants. The $\chi_{ab}$ matrix, along with the overtone and combination band frequencies and
intensities obtained from our computational protocol is given in the Supporting Information.

In the coupled anharmonic oscillator model, vibrational energy levels are generated by
exact state counting using a breadth-first search algorithm over the
integer quantum number lattice \cite{chen2026rrao_anharmonic}. The search
starts from the zero-point state and recursively adds single-quantum
excitations, retaining all states whose anharmonic excitation energies
lie below the maximum energy required for the master equation grid. This avoids using a separable density of states approximation and keeps the off-diagonal anharmonic couplings in the state counting.

Let $\mathcal B$ be the set of anharmonic states generated by this search. We define the bin indicator
\begin{equation}
\Theta_i(\mathbf n)
=
\begin{cases}
1, & E_i-\Delta E/2 \le E(\mathbf n) < E_i+\Delta E/2,\\
0, & \mathrm{otherwise},
\end{cases}
\label{eq:bin-indicator}
\end{equation}
and the number of states in bin $i$ as
\begin{equation}
\Omega_i=\sum_{\mathbf n\in\mathcal B}\Theta_i(\mathbf n).
\label{eq:bin-state-count}
\end{equation}
The microcanonical probability that mode $a$ contains $q$ quanta in
energy grain $i$ is then
\begin{equation}
P_i(n_a=q)
=
\frac{1}{\Omega_i}
\sum_{\mathbf n\in\mathcal B}
\Theta_i(\mathbf n)\,\delta_{n_a,q}.
\label{eq:microcanonical-probability}
\end{equation}
For a combination band involving modes $a$ and $b$, the corresponding
joint probability is
\begin{equation}
P_i(n_a=q,n_b=r)
=
\frac{1}{\Omega_i}
\sum_{\mathbf n\in\mathcal B}
\Theta_i(\mathbf n)\,
\delta_{n_a,q}\delta_{n_b,r}.
\label{eq:joint-probability}
\end{equation}
These probabilities are normalized within each energy grain and provide
the microcanonical weights for radiative transition strengths.

Radiative transitions are taken from the VPT2 transition list and include fundamentals, first overtones, and $1+1$ combination bands. For a transition channel $\ell$ with quantum-number step $\mathbf d_\ell$,
the state-dependent absorption and emission wavenumbers are
\begin{equation}
\nu_\ell^{+}(\mathbf n)
=
E(\mathbf n+\mathbf d_\ell)-E(\mathbf n),
\label{eq:transition-frequency-up}
\end{equation}
and
\begin{equation}
\nu_\ell^{-}(\mathbf n)
=
E(\mathbf n)-E(\mathbf n-\mathbf d_\ell).
\label{eq:transition-frequency-down}
\end{equation}
Here $\mathbf d_\ell=\mathbf e_a$ for a fundamental transition,
$\mathbf d_\ell=2\mathbf e_a$ for a first overtone, and
$\mathbf d_\ell=\mathbf e_a+\mathbf e_b$ for a $1+1$ combination band.

\begin{table}[t]
\centering
\caption{PBE0-D4/def2-TZVPD VPT2 vibrational frequencies and intensities for
\ch{(H2O)2Cl-}. Here $\omega_a$ is the harmonic frequency,
$\nu_a^{1,0}$ is the anharmonic fundamental transition frequency,
$\chi_{aa}$ is the diagonal anharmonicity, and $I_{1,0}$ is the
fundamental IR intensity. The full anharmonicity matrix, together with
the overtone and combination band frequencies and intensities used in
the master equation calculations, is given in the Supporting
Information.}
\label{tab:h2o2cl_pbe0d4_fundamentals}
\renewcommand{\arraystretch}{1.15}
\setlength{\tabcolsep}{4.5pt}
\begin{tabular}{r r r r r}
\hline
Mode &
$\omega_a$ &
$\nu_a^{1,0}$ &
$\chi_{aa}$ &
$I_{1,0}$ \\
&
(cm$^{-1}$) &
(cm$^{-1}$) &
(cm$^{-1}$) &
(km mol$^{-1}$) \\
\hline
 1 &   99.203 &   57.182 &   -0.995 &    0.477 \\
 2 &  159.946 &  133.920 &   -3.243 &    4.877 \\
 3 &  189.424 &  140.640 &   -8.348 &   89.189 \\
 4 &  214.839 &  209.613 &   -1.665 &   31.448 \\
 5 &  354.385 &  339.837 &  -12.615 &   23.904 \\
 6 &  416.392 &  376.824 &   -1.462 &   40.772 \\
 7 &  482.543 &  394.082 &  -22.986 &   15.488 \\
 8 &  681.325 &  624.925 &  -23.202 &   96.722 \\
 9 &  819.441 &  750.250 &  -60.246 &  106.616 \\
10 & 1661.990 & 1696.774 &    3.204 &   86.798 \\
11 & 1701.930 & 1667.994 &  -16.948 &  178.834 \\
12 & 3280.065 & 2977.094 & -176.825 & 1003.730 \\
13 & 3597.023 & 3325.093 & -143.082 &  512.817 \\
14 & 3786.978 & 3629.283 & -104.782 &  141.857 \\
15 & 3898.827 & 3720.946 &  -80.610 &   27.450 \\
\hline
\end{tabular}
\end{table}

The hot band transition strengths are scaled following the anharmonic cascade
model of infrared cooling \cite{
stockett2025infraredcooling}. Each VPT2 fundamental, overtone, or
combination band intensity is used as an anchor intensity. The
state-dependent Einstein coefficient is approximated as \cite{brenner1993population,
stockett2025infraredcooling}
\begin{equation}
A_\ell^{\pm}(\mathbf n)
=
S_\ell^{\pm}(\mathbf n)\,
A_\ell^0
\left[
\frac{\nu_\ell^{\pm}(\mathbf n)}
{\nu_\ell^0}
\right]^3 ,
\label{eq:einstein-scaling-free}
\end{equation}
where $A_\ell^0$ and $\nu_\ell^0$ are the reference Einstein coefficient
and transition wavenumber from the VPT2 list. The harmonic
occupation-number factors are \cite{brenner1993population,
stockett2025infraredcooling}
\begin{equation}
\begin{aligned}
S_a^{+}&=n_a+1,
&\qquad S_a^{-}&=n_a,\\
S_{2a}^{+}&=\frac{(n_a+1)(n_a+2)}{2},
& S_{2a}^{-}&=\frac{n_a(n_a-1)}{2},\\
S_{ab}^{+}&=(n_a+1)(n_b+1),
& S_{ab}^{-}&=n_a n_b .
\end{aligned}
\label{eq:scaling-factors}
\end{equation}
The upward and downward rates are then weighted by the thermal radiation field,
\begin{equation}
k_\ell^{\rm abs}(\mathbf n)
=
A_\ell^{+}(\mathbf n)\,
\bar n_T[\nu_\ell^{+}(\mathbf n)],
\label{eq:abs-rate-free}
\end{equation}
and
\begin{equation}
k_\ell^{\rm em}(\mathbf n)
=
A_\ell^{-}(\mathbf n)
\left\{1+\bar n_T[\nu_\ell^{-}(\mathbf n)]\right\}.
\label{eq:em-rate-free}
\end{equation}
Thus absorption is proportional to the thermal photon occupation
\begin{equation}
\bar n_T(\nu)=
\left[
\exp\left(\frac{hc\nu}{k_{\rm B}T}\right)-1
\right]^{-1},
\label{eq:planck-occupation}
\end{equation}
whereas emission contains both spontaneous and stimulated contributions. In this
section, all radiative rates are defined in free space. The modification
of these rates by a structured electromagnetic environment will be introduced separately in the following section.

The coarse-grained contribution of transition channel $\ell$ to the
rate from bin $i$ to bin $j$ is
\begin{equation}
k_{i\to j}^{\ell,\pm}
=
\frac{1}{\Omega_i}
\sum_{\mathbf n\in\mathcal B}
\Theta_i(\mathbf n)\,
\Theta_j(\mathbf n\pm\mathbf d_\ell)\,
k_\ell^{\pm}(\mathbf n).
\label{eq:coarse-transition-rate}
\end{equation}
The total rate $k_{i\to j}$ is obtained by summing
Eq.~\eqref{eq:coarse-transition-rate} over all allowed absorption and
emission channels. Because the transition frequencies depend on the coupled anharmonic state, different hot band, overtone, and combination band transitions originating from the same energy grain may populate different final grains.

Fundamental frequencies, IR intensities, and diagonal anharmonicities
used for \((\mathrm{H_2O})_2\mathrm{Cl}^-\) are summarized in
Table~\ref{tab:h2o2cl_pbe0d4_fundamentals}.
For \(E_0=3500~\mathrm{cm^{-1}}\)
(\(10.01~\mathrm{kcal~mol^{-1}}\)), the free-space calculation at
\(T=300~\mathrm{K}\) gives
\(k_{\rm BIRD}=2.4\times10^{-3}~\mathrm{s^{-1}}\), reproducing the
experimental BIRD rate reported by Dunbar et al. and remaining
consistent with the \(10.1~\mathrm{kcal~mol^{-1}}\) threshold obtained
in their harmonic master-equation simulation \cite{dunbar1995zero}. These results validate our microscopic model. Therefore, for the microcavity and bath-gas calculations below, we use this \(E_0\) value and the corresponding free-space rate as the reference for
computing relative BIRD rates in different electromagnetic and collisional environments.

\iffalse

\begin{figure}[t]
\centering
\includegraphics[width=\columnwidth]{figures/paper_bird_rate_vs_E0_APS.png}
\caption{Free-space BIRD dissociation rate of \ch{(H2O)2Cl-} as a
function of the dissociation threshold $E_0$ at $T=300~\mathrm{K}$.
The coupled anharmonic master equation calculation includes
fundamental, first overtone, and $1+1$ combination band transitions.
The dashed horizontal line indicates the experimental rate
$k_{\rm BIRD}^{\rm exp}=2.4\times10^{-3}~\mathrm{s^{-1}}$ reported by
Dunbar et al.~\cite{dunbar1995zero}. Inclusion of overtone and
combination band pathways substantially changes the BIRD rates, shifting the inferred threshold energy
relative to a fundamentals only description.}
\label{fig:h2o2cl_e0_anharmonic_benchmark}
\end{figure}

\fi

\subsection{Radiative transition rates in planar cavities}
\label{sec:multilayer_rates}

In the thermal radiative dissociation mechanism considered here, the reactant remains weakly coupled
to the electromagnetic field, so the molecular states are not hybridized with microcavity modes. The resonator changes the radiative transition rates
only via the electromagnetic density of states entering the
Einstein-coefficient description \cite{fermi1932quantum,
einsteinradiation,suyabatmaz2025polaritonic,barnes_classical_2020}.
For a transition $i\rightarrow j$ with frequency
$\omega_{ij}$, the free-space spontaneous emission rate is
\begin{equation}
A_{ij}^{(0)}
=
\frac{\pi\omega_{ij}|\mu_{ij}|^2}
{3\epsilon_0\hbar}
D_0(\omega_{ij}),
\qquad
D_0(\omega)=\frac{\omega^2}{\pi^2c^3}.
\label{eq:Aij-free}
\end{equation}

Inside a resonator the free-space density of states is replaced by the
isotropically averaged microcavity density of states $D_C(\omega)$. We define the DOS enhancement factor
\begin{equation}
F_{\rm env}(\omega)
=
\frac{D_C(\omega)}{D_0(\omega)} .
\label{eq:Fenv_def}
\end{equation}

The spontaneous emission coefficient is therefore renormalized as
$A_{ij}^{C}=A_{ij}^{(0)}F_{\rm env}(\omega_{ij})$. The 
radiation energy density inside the microcavity is modified in the same way,
$\rho_C(\omega,T)=\hbar\omega\,n_{\rm BE}(\omega,T)D_C(\omega)
=F_{\rm env}(\omega)\rho_0(\omega,T)$. Consequently, the absorption
rate for the upward transition $j\rightarrow i$ and the stimulated
emission rate for the downward transition $i\rightarrow j$ are scaled as
$k_{ji}^{\rm abs,C}
=F_{\rm env}(\omega_{ij})k_{ji}^{\rm abs,(0)}$ and
$k_{ij}^{\rm stim,C}
=F_{\rm env}(\omega_{ij})k_{ij}^{\rm stim,(0)}$, respectively.

\begin{figure}[t]
    \centering
    \includegraphics[width=\columnwidth]{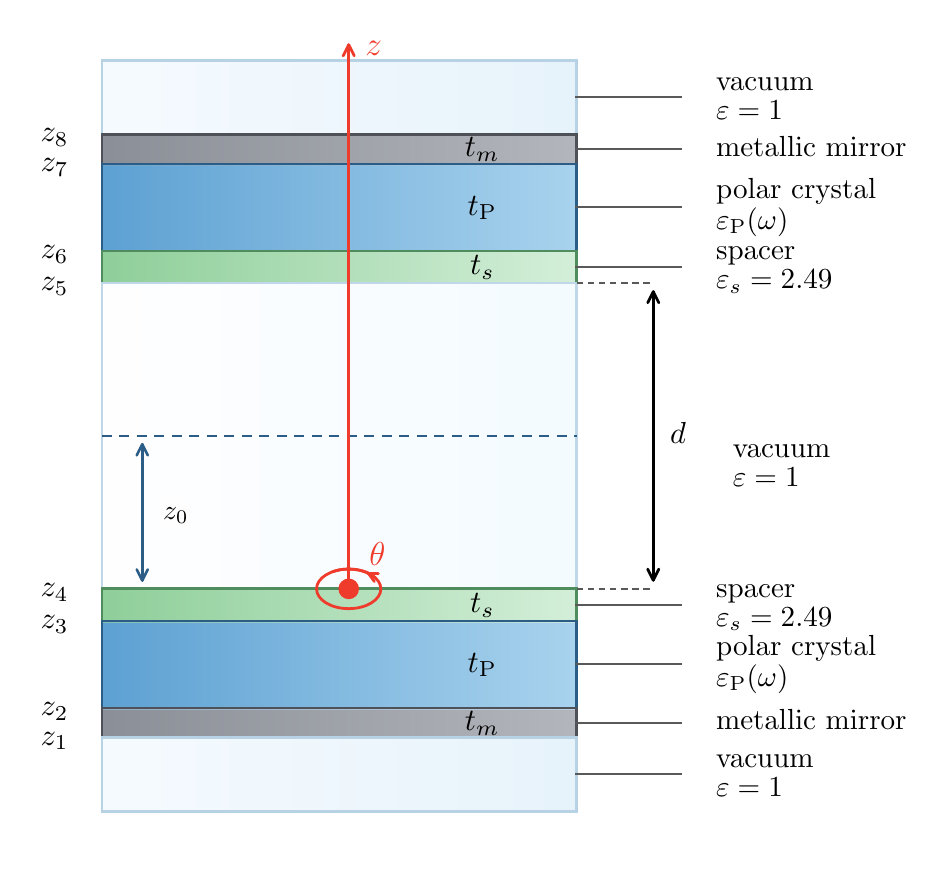}
    \caption{Planar multilayer cavity used to compute the photonic LDOS in the vacuum reaction region. Two metallic mirrors of thickness $t_m$ confine the field. Each mirror is backed by a semi-infinite vacuum region. Inside the cavity, polar crystal layers of thickness $t_P$ with permittivity $\varepsilon_P(\omega)$ are separated from the central vacuum gap by dielectric spacers of thickness $t_s$ with $\varepsilon_s=2.49$.}
    \label{fig:multilayer_schematic}
\end{figure}

In our earlier study we considered diatomic infrared radiative dissociation in conventional metallic microcavities \cite{suyabatmaz2025polaritonic}. Here, the structured environment considered is the planar multilayer
cavity shown in Fig.~\ref{fig:multilayer_schematic}. Two metallic
mirrors confine the electromagnetic field. Inside
the cavity, two polar crystal layers are separated from the central vacuum reaction region by thin dielectric spacers. The reactant molecule is placed in the vacuum gap where it rotates freely, so we use the
isotropically averaged cavity DOS, $D_C(\omega)$, in that region. Even though the molecule sits in vacuum, it experiences a strongly modified electromagnetic density of states due to the surface phonon polariton
and microcavity modes supported by the surrounding polar crystal layers
\cite{joulain2005surface,novotny_hecht_2012}.

We compute the microcavity LDOS $D_{C}^{L}(\omega)$ from the dyadic Green tensor of
Maxwell's equations for a planar stratified medium \cite{wylie_sipe_1984,
tomas1995green,chew1995waves,novotny_hecht_2012}. With the convention
used in Eq.~\eqref{eq:Aij-free}, the isotropic LDOS is
\begin{equation}
D_{C}^{L}(\omega)(\omega,\mathbf r)
=
\frac{2\omega}{\pi c^2}
\Im\Tr\,\mathbf G(\mathbf r,\mathbf r;\omega),
\label{eq:electric_ldos_trace}
\end{equation}
where $\mathbf G$ is the electric Green tensor. In a planar geometry the in-plane wavevector $k_\parallel$ is conserved.
For a layer with permittivity $\varepsilon(\omega)$,
\begin{equation}
k_z(\omega,k_\parallel)
=
\left[
\varepsilon(\omega)\frac{\omega^2}{c^2}
-
k_\parallel^2
\right]^{1/2},
\qquad
\Im k_z\ge 0 .
\label{eq:kz_def}
\end{equation}
In the vacuum reaction region, \(k_0=\omega/c\). Modes with
$0\le k_\parallel\le k_0$ are propagating, whereas
$k_\parallel>k_0$ gives $k_z=i\kappa, ~\kappa \in \mathbb{R}$ with
\begin{equation}
\kappa(\omega,k_\parallel)
=
\left(k_\parallel^2-k_0^2\right)^{1/2},
\qquad
\Re\kappa\ge 0 .
\label{eq:kappa_def}
\end{equation}
The latter are evanescent waves whose fields decay away from the interfaces over a length scale set by $1/\Re\kappa$ \cite{joulain2005surface,
novotny_hecht_2012}.

Reflection from the multilayer stacks above and below the observation point is described by effective TE ($s$) and TM ($p$) reflection amplitudes. We denote by $r_\uparrow^{s,p}(\omega,k_\parallel)$ the reflection coefficient seen from within the vacuum gap looking upward and by
$r_\downarrow^{s,p}(\omega,k_\parallel)$ the reflection coefficient seen looking downward  \cite{tomas1995green,chew1995waves}. The Fabry-Pérot
denominators that sum all round trips across a vacuum gap of thickness
$d$ are
\begin{equation}
C_{s,p}
=
1
-
r_\uparrow^{s,p}
r_\downarrow^{s,p}
e^{2ik_z d}.
\label{eq:fp_denominator}
\end{equation}
Here and below the arguments $(\omega,k_\parallel)$ are omitted for convenience. We evaluate the LDOS at positions $z$ inside
the vacuum gap, measured from the lower interface, and then average over
the central reaction region, i.e.,
\begin{equation}
{D}_{C}(\omega)
=
\frac{1}{d}\int_0^d
D_{C}^{L}(\omega,z)\,dz .
\label{eq:z_averaged_ldos}
\end{equation}

Using the TE and TM Weyl expansion for planar multilayers, the
scattering contributions to the coincident-point Green tensor can be
written in terms of the effective reflection amplitudes and the
round-trip denominators (Eq. \ref{eq:fp_denominator}) \cite{wylie_sipe_1984,tomas1995green,
chew1995waves,novotny_hecht_2012}. We define
\begin{equation}
A_{s,p}(z)
=
r_\uparrow^{s,p}e^{2ik_z z}
+
r_\downarrow^{s,p}e^{2ik_z(d-z)}
\label{eq:Asp_def}
\end{equation}
and
\begin{equation}
B_{s,p}
=
2r_\uparrow^{s,p}r_\downarrow^{s,p}e^{2ik_zd},
\label{eq:Bsp_def}
\end{equation}
and microcavity response factors
\begin{equation}
R_{s,p}(\omega,k_\parallel;z)
=
\frac{A_{s,p}(z)+B_{s,p}}{C_{s,p}} .
\label{eq:Rsp_def}
\end{equation}

With these definitions, the imaginary parts of the relevant Green-tensor
components in the vacuum gap are
\begin{equation}
\begin{split}
\Im G_{zz}(\omega;z)
&=
\frac{k_0}{6\pi}
+
\frac{1}{4\pi}
\Re\int_0^\infty
\frac{k_\parallel^3}
{k_0^2 k_z}
R_p
\,dk_\parallel ,
\end{split}
\label{eq:ImGzz_fullint}
\end{equation}
and
\begin{equation}
\begin{split}
\Im G_{xx}(\omega;z)
&=
\frac{k_0}{6\pi}
+
\frac{1}{8\pi}
\Re\int_0^\infty
\frac{k_\parallel}{k_z}
\\
&\quad\times
\left[
R_s
-
\frac{k_\parallel^2}{k_0^2}R_p
\right]
dk_\parallel .
\end{split}
\label{eq:ImGxx_fullint}
\end{equation}
The trace entering Eq.~\eqref{eq:electric_ldos_trace} is
\begin{equation}
\Im\Tr\,\mathbf G
=
2\Im G_{xx}+\Im G_{zz},
\label{eq:green_trace_planar}
\end{equation}
where symmetry gives $G_{xx}=G_{yy}$.

The $k_\parallel$ integrals in
Eqs.~\eqref{eq:ImGzz_fullint}-\eqref{eq:ImGxx_fullint} naturally
separate into propagating and evanescent contributions,
\begin{equation}
\int_0^\infty dk_\parallel
=
\int_0^{k_0} dk_\parallel
+
\int_{k_0}^{\infty} dk_\parallel .
\label{eq:prop_evan_split}
\end{equation}
The first term gives the propagating LDOS, while the second term gives
the evanescent LDOS. Because the molecular position is averaged over the vacuum
reaction region and because evanescent fields decay away from the
interfaces, the evanescent contribution is most important for narrow
cavities and for transitions near surface phonon polariton resonances of
the polar layers \cite{joulain2005surface,novotny_hecht_2012}.

The metallic mirrors are modeled with a Drude permittivity
\cite{drude1900,rakic1998optical,barnes_classical_2020}
\begin{equation}
\varepsilon_m(\omega)
=
\varepsilon_{\infty,m}
-
\frac{\omega_{p,m}^2}
{\omega^2+i\gamma_m\omega}.
\label{eq:drude_mirror}
\end{equation}
Here $\varepsilon_{\infty,m}$ is the high-frequency background
permittivity, $\omega_{p,m}$ is the plasma frequency, and $\gamma_m$ is
the electronic damping rate. For the Au mirrors, we use
$\varepsilon_{\infty,m}=1.0$, $\omega_{p,m}=1.22\times10^{16}~
\mathrm{rad\,s^{-1}}$, and $\gamma_m=7.11\times10^{13}~
\mathrm{rad\,s^{-1}}$\cite{rakic1998optical,barnes_classical_2020}.

The polar crystal layers are described by a Lorentz phonon dielectric
function \cite{born_wolf_1999,komandin2009mgo}
\begin{equation}
\varepsilon_P(\omega)
=
\varepsilon_{\infty,P}
\frac{
\omega^2-\omega_{\rm LO}^2+i\gamma_{\rm LO}\omega
}{
\omega^2-\omega_{\rm TO}^2+i\gamma_{\rm TO}\omega
}.
\label{eq:lorentz_polar}
\end{equation}
Here $\varepsilon_{\infty,P}$ is the high-frequency dielectric constant
of the polar crystal, $\omega_{\rm TO}$ and $\omega_{\rm LO}$ are the
transverse and longitudinal optical phonon frequencies, and
$\gamma_{\rm TO}$ and $\gamma_{\rm LO}$ are the corresponding phonon
damping rates. For MgO, we use parameters from infrared/terahertz
dispersion measurements of single-crystal MgO from Ref. \cite{komandin2009mgo} where 
$\varepsilon_{\infty,P}=2.95$, $\omega_{\rm TO}=396~\mathrm{cm^{-1}}$,
$\omega_{\rm LO}=720~\mathrm{cm^{-1}}$, $\gamma_{\rm TO}=6.2~
\mathrm{cm^{-1}}$, and $\gamma_{\rm LO}=13.0~\mathrm{cm^{-1}}$.
The dielectric spacers separating the MgO layers from the central
vacuum reaction region are taken to have
$\varepsilon_s=2.49$ \cite{barnes_classical_2020}.

\subsection{Collisional activation and pressure dependence}
\label{sec:methods_pressure}

To account for energy transfer in collisions with a background gas, we
extended our zero-pressure master equation of Sec.~IIA by adding a collisional transport operator \cite{gilbert_smith_1990,barker2001multiwell,
glowacki2012mesmer,klippenstein2022spiers,dunbar2004birdreview}. The
population vector over energy grains, $\boldsymbol{p}(t)$, then evolves
as
\begin{equation}
\frac{d\boldsymbol{p}}{dt}
= -
\left[
\mathbf{J}_{\rm rad}(L)
+
\mathbf{J}_{\rm coll}(T,P)
+
\mathbf{J}_{\rm diss}
\right]
\boldsymbol{p}.
\label{eq:pressure_master_equation}
\end{equation}
Here $\mathbf{J}_{\rm rad}(L)$ is the microcavity-modified radiative
transport operator defined above, $\mathbf{J}_{\rm coll}(T,P)$ describes
redistributive collisional energy transfer among grains, and
$\mathbf{J}_{\rm diss}$ is the diagonal loss matrix corresponding to dissociation.
%In the absence of reaction,
%$\mathbf{J}_{\rm rad}$ and $\mathbf{J}_{\rm coll}$ conserve %total population.

We estimate collisional energy transfer with a single-collision kernel
$P_{\rm coll}(i\leftarrow j)$ defined on the same energy grid
\cite{gilbert_smith_1990,barker2001multiwell,glowacki2012mesmer}. For a
downward step from grain $j$ with energy $E_j$ to grain $i<j$ with
energy $E_i$, we use the standard exponential-down model
\cite{tardy1977ivet,gilbert_smith_1990,barker2001multiwell,
jasper2009collisional,klippenstein2022spiers}
\begin{equation}
P_\downarrow(i\leftarrow j)
\propto
\exp\left[
-\frac{E_j-E_i}
{\langle\Delta E_{\rm down}\rangle}
\right],
\label{eq:exp_down_kernel}
\end{equation}
with an average downward step size
$\langle\Delta E_{\rm down}\rangle\simeq 100~\mathrm{cm^{-1}}$.
Upward steps are obtained by imposing detailed balance with respect to the discretized microcanonical density of states $g_i$ and the Boltzmann factor at the bath temperature \cite{gilbert_smith_1990,barker2001multiwell,
glowacki2012mesmer}
\begin{equation}
\frac{P_{\rm coll}(i\leftarrow j)}
{P_{\rm coll}(j\leftarrow i)}
=
\frac{g_i}{g_j}
\exp\left[
-\frac{E_i-E_j}{k_{\rm B}T}
\right],
\label{eq:coll_detailed_balance}
\end{equation}
and each column of $P_{\rm coll}$ is normalized so that
$\sum_i P_{\rm coll}(i\leftarrow j)=1$. The DOS weights $g_i$ are
obtained by exact combinatorial state counting over the vibrational
manifold and summing over all microstates in grain $i$
\cite{beyer_swinehart1973,stein_rabinovitch1973}. Further details of the
kernel construction are given in the Supplementary Information.

The collisional transport operator is written in terms of the
per-molecule collision frequency $Z(T,P)$ as
 \cite{gilbert_smith_1990,barker2001multiwell,glowacki2012mesmer,
cho2023collisional}
\begin{equation}
\left(\mathbf{J}_{\rm coll}\right)_{ij}
=
\begin{cases}
Z(T,P)\,P_{\rm coll}(i\leftarrow j),
& i\neq j, \\[4pt]
-Z(T,P)\displaystyle\sum_{k\neq i}
P_{\rm coll}(k\leftarrow i),
& i=j .
\end{cases}
\label{eq:Kcoll_definition}
\end{equation}
Thus each column of $\mathbf{J}_{\rm coll}$ sums to zero, ensuring collisions conserve the total population.

We evaluate $Z(T,P)$ microscopically using kinetic theory
for collisions between the ion and a methane bath gas
\cite{chapman_cowling_1970,hirschfelder1954mtgl,neufeld1972collision}
The bimolecular collision rate coefficient $k_{\rm coll}(T)$ is estimated from the Lennard-Jones parameters and the Chapman-Enskog collision integral, and the collision frequency is
\cite{chapman_cowling_1970,hirschfelder1954mtgl,neufeld1972collision}
\begin{align}
&Z(T,P)
=
k_{\rm coll}(T)\,n_{\rm bath}(T,P), \nonumber \\ 
& n_{\rm bath}(T,P)
=
\frac{x_{\rm bath}P}{k_{\rm B}T},
\label{eq:collision_frequency}
\end{align}
where $x_{\rm bath}$ is the bath-gas mole fraction.

Molecule-wall collisions are not included in the present gas-phase
pressure-dependent model. This approximation follows the experimental setup in standard BIRD measurements, where gas-phase ions are stored in Fourier-transform ion cyclotron resonance (FT-ICR) or related
ion-trapping mass spectrometers under high-vacuum conditions and are
activated by the ambient blackbody radiation field rather than by direct wall impacts \cite{dunbar2004birdreview,price1996bird,
schnier1996blackbody}. Price et al. reported
BIRD measurements for ions stored at pressures below $10^{-8}~\mathrm{Torr}$,
with dissociation rate constants becoming independent of pressure below
approximately $10^{-7}~\mathrm{Torr}$, consistent with activation by
blackbody photons emitted by the chamber walls rather than by collisional
heating \cite{price1996bird}. The same zero-pressure radiative-activation
picture has been used for hydrated chloride cluster ions, including
$(\mathrm{H_2O})_2\mathrm{Cl}^{-}$ and
$(\mathrm{H_2O})_3\mathrm{Cl}^{-}$, where the measured unimolecular decay was
modeled as thermal radiation induced dissociation of isolated trapped ions
\cite{dunbar1995zero,dunbar1998ambient}. Accordingly, our model
assumes that the molecular ion remains in the central vacuum region, so
that the resonator boundaries modify the electromagnetic local density of
states sampled by the ion, while direct molecule-wall collisions are
excluded. Background gas collisions are included separately through the
pressure-dependent collisional activation operator.

The pressure-dependent dissociation rate $k_{\rm diss}(T,P,L)$ is
obtained from the slowest decaying eigenmode of the transport generator in
Eq.~\eqref{eq:pressure_master_equation}, using the same absorbing-boundary
condition as in the zero-pressure BIRD model
\cite{vankampen2007,gardiner2009,dunbar1995zero,dunbar1998ambient,
dunbar2004birdreview,salzburger2024mem}. A detailed description of the collision model and parameter choices is provided in the Supplementary
Information.

\section{Results and Discussion}
\subsection{Microcavity-assisted thermal infrared radiative dissociation} 

We considered first the case of dissociation at zero-pressure conditions in an Au/MgO multilayer structure shown in
Fig.~\ref{fig:multilayer_schematic}. The total microcavity length \(L\)
denotes the distance between the two Au mirrors across the internal
multilayer region. The Au layers have thickness
\(t_{\rm Au}=100~\mathrm{nm}\) on each side. Each setup contains two
MgO layers of thickness \(t_{\rm MgO}\), two dielectric spacer layers
of thickness \(t_s=0.2~\mu\mathrm{m}\) with \(\varepsilon_s=2.49\),
and a central vacuum reaction gap of thickness
\(d=L-2t_{\rm MgO}-2t_s\), where the reactive cluster is placed. We
consider 70 Au/MgO geometries obtained from five MgO thicknesses,
\(t_{\rm MgO}=0.5,1.0,1.5,2.0,\) and \(2.5~\mu\mathrm{m}\), and
fourteen cavity lengths from \(L=5.5\) to \(10~\mu\mathrm{m}\). The Au
mirrors are modeled with a Drude response using
\(\varepsilon_{\infty}=1.0\), \(\omega_p=1.22\times10^{16}~
\mathrm{rad\,s^{-1}}\), and \(\gamma=7.11\times10^{13}~
\mathrm{rad\,s^{-1}}\) \cite{rakic1998optical,barnes_classical_2020}.
The MgO layers are described by a Lorentz phonon model with
\(\varepsilon_{\infty}=2.95\), \(\omega_{\rm TO}=396~\mathrm{cm^{-1}}\),
\(\omega_{\rm LO}=720~\mathrm{cm^{-1}}\),
\(\gamma_{\rm TO}=6.2~\mathrm{cm^{-1}}\), and
\(\gamma_{\rm LO}=13.0~\mathrm{cm^{-1}}\), taken from Ref. \cite{komandin2009mgo}.
Additional computations for other metallic mirrors, including Al and Pt, and for SiC as an alternative polar crystal are provided in the
Supporting Information.

\begin{figure}[t]
    \centering
    \includegraphics[width=\columnwidth]{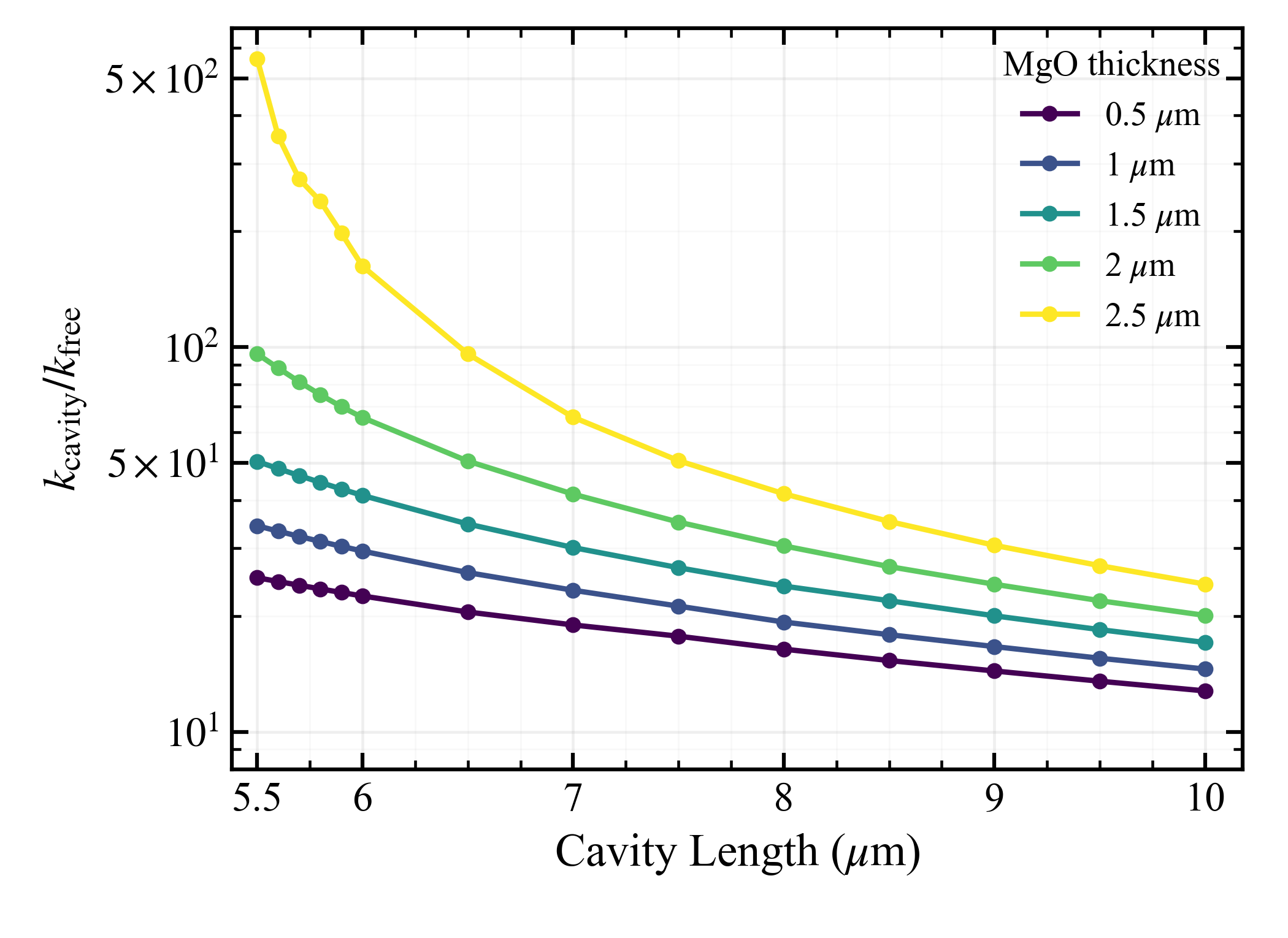}
    \caption{Microavity-enhanced BIRD rate for selected Au/MgO multilayer
    geometries, shown as \(k_{\rm cavity}/k_{\rm free}\) versus cavity
    length. The color indicates the MgO layer thickness. Short cavities
    and thicker MgO layers produce the largest enhancements because the
    vacuum reaction region samples a stronger evanescent near-field
    contribution from the polar interfaces.}
    \label{fig:mgo_rate_70_geometries}
\end{figure}

Figure~\ref{fig:mgo_rate_70_geometries} summarizes the microcavity-length dependence of the BIRD rate across the considered Au/MgO geometries. The rate is enhanced for every geometry considered, but the enhancement is strongly geometry dependent. For a fixed MgO thickness, increasing the distance between metallic mirrors  monotonically decreases \(k_{\rm cavity}/k_{\rm free}\).
For \(t_{\rm MgO}=0.5~\mu\mathrm{m}\), the rate enhancement decreases
from \(25.2\) at \(L=5.5~\mu\mathrm{m}\) to \(12.8\) at
\(L=10~\mu\mathrm{m}\). For \(t_{\rm MgO}=2.5~\mu\mathrm{m}\), the
enhancement decreases from \(562\) to \(24.2\) over the same length
range. Thus, the largest rate enhancements occur for thick MgO layers
and short cavities, where the central vacuum reaction region is narrow
and the reactive cluster samples the strongest evanescent near field
from the polar interfaces.

The DOS and dispersion plots clarify the physical origin of this trend. The enhancement is concentrated in and near the MgO Reststrahlen band. For the Lorentz phonon response, the Reststrahlen band is the interval
\begin{equation}
\omega_{\rm TO}<\omega<\omega_{\rm LO},
\qquad
\mathrm{Re}\,\varepsilon_{\rm MgO}(\omega)<0 .
\label{eq:reststrahlen_condition}
\end{equation}
This frequency window supports surface
phonon-polariton (SPhP) modes localized near the MgO interfaces. In the
nonretarded single interface limit, the SPhP pole is determined by
\begin{equation}
\varepsilon_{\rm MgO}(\omega_s)+\varepsilon_d=0 ,
\label{eq:sphp_pole_condition}
\end{equation}
where \(\varepsilon_d\) is the permittivity of the adjacent medium.
Equivalently, in the lossless limit,
\begin{equation}
\omega_s^2
\simeq
\frac{
\varepsilon_\infty\omega_{\rm LO}^2
+\varepsilon_d\omega_{\rm TO}^2
}{
\varepsilon_\infty+\varepsilon_d
}.
\label{eq:sphp_frequency_estimate}
\end{equation}
Using \(\varepsilon_d=2.49\) for the spacer and \(\varepsilon_d=1\) for vacuum gives lossy poles
\(\omega_s^{\rm MgO/spacer}\simeq 594.0-i4.9~\mathrm{cm^{-1}}\) and
\(\omega_s^{\rm MgO/vac}\simeq 653.3-i5.6~\mathrm{cm^{-1}}\). The
associated linewidth is given by
\(\Gamma_s=-2\,\mathrm{Im}\,\omega_s\), giving approximately \(9.8\)
and \(11.2~\mathrm{cm^{-1}}\), respectively. In the complete Au/MgO
multilayer these resonances are broadened, split, and hybridized by the finite MgO thickness, spacer layers, Au mirrors, and central vacuum gap. This hybridized SPhP response produces the large DOS peak and the broadband DOS enhancement inside the Reststrahlen band.

\begin{figure}[t]
    \centering
    \includegraphics[width=\columnwidth]{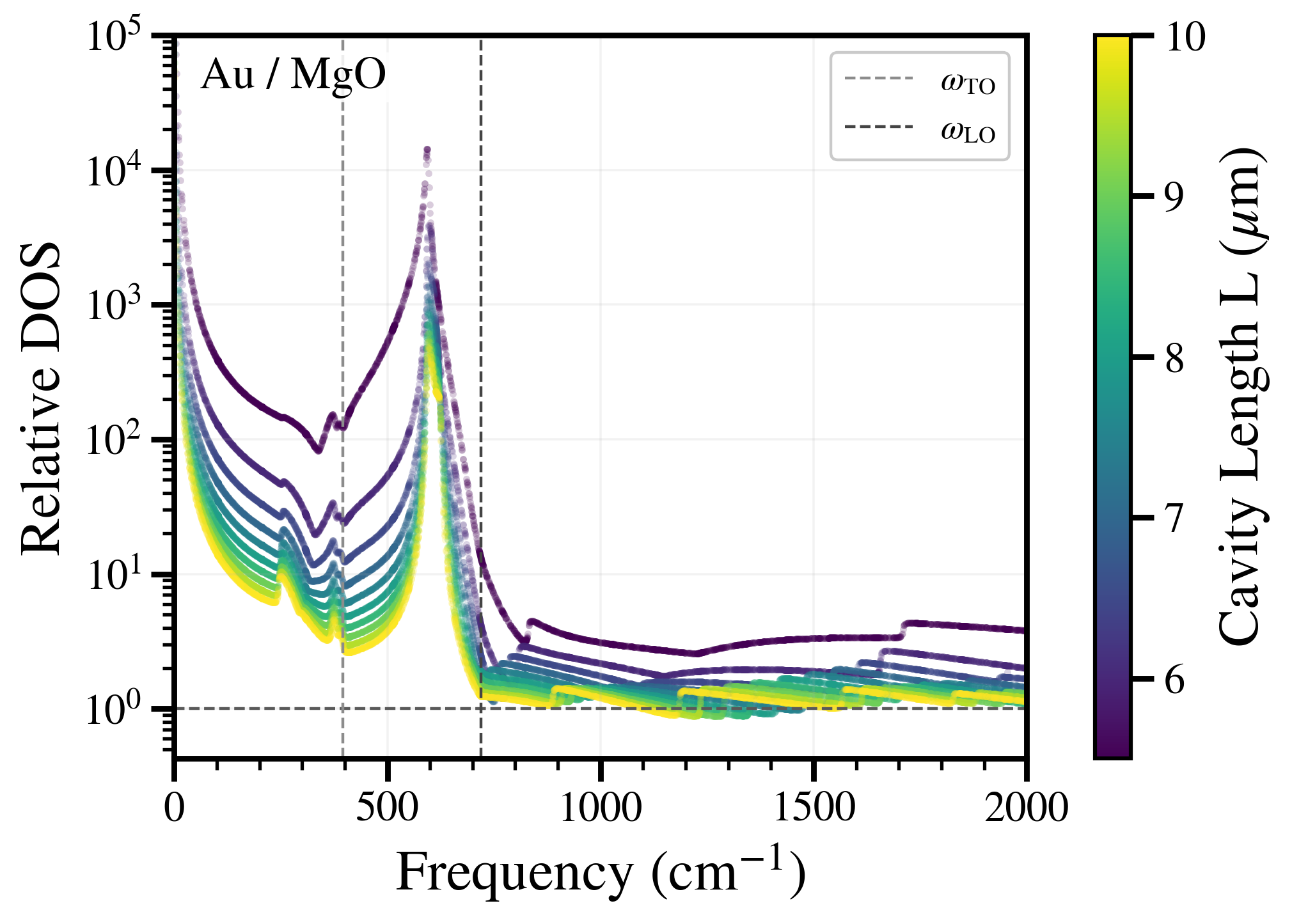}
    \caption{Relative photonic DOS, \(D_C/D_0\), in the vacuum
    reaction region of the Au/MgO multilayer cavity for a fixed MgO
    thickness of \(2.5~\mu\mathrm{m}\) and varying cavity length. The
    vertical dashed lines indicate \(\omega_{\rm TO}\) and
    \(\omega_{\rm LO}\) of MgO. The strongest DOS enhancement occurs
    inside the MgO Reststrahlen band, where the polar layers support
    hybridized SPhP modes.}
    \label{fig:mgo_dos}
\end{figure}

\begin{figure}[t]
    \centering
    \includegraphics[width=\columnwidth]{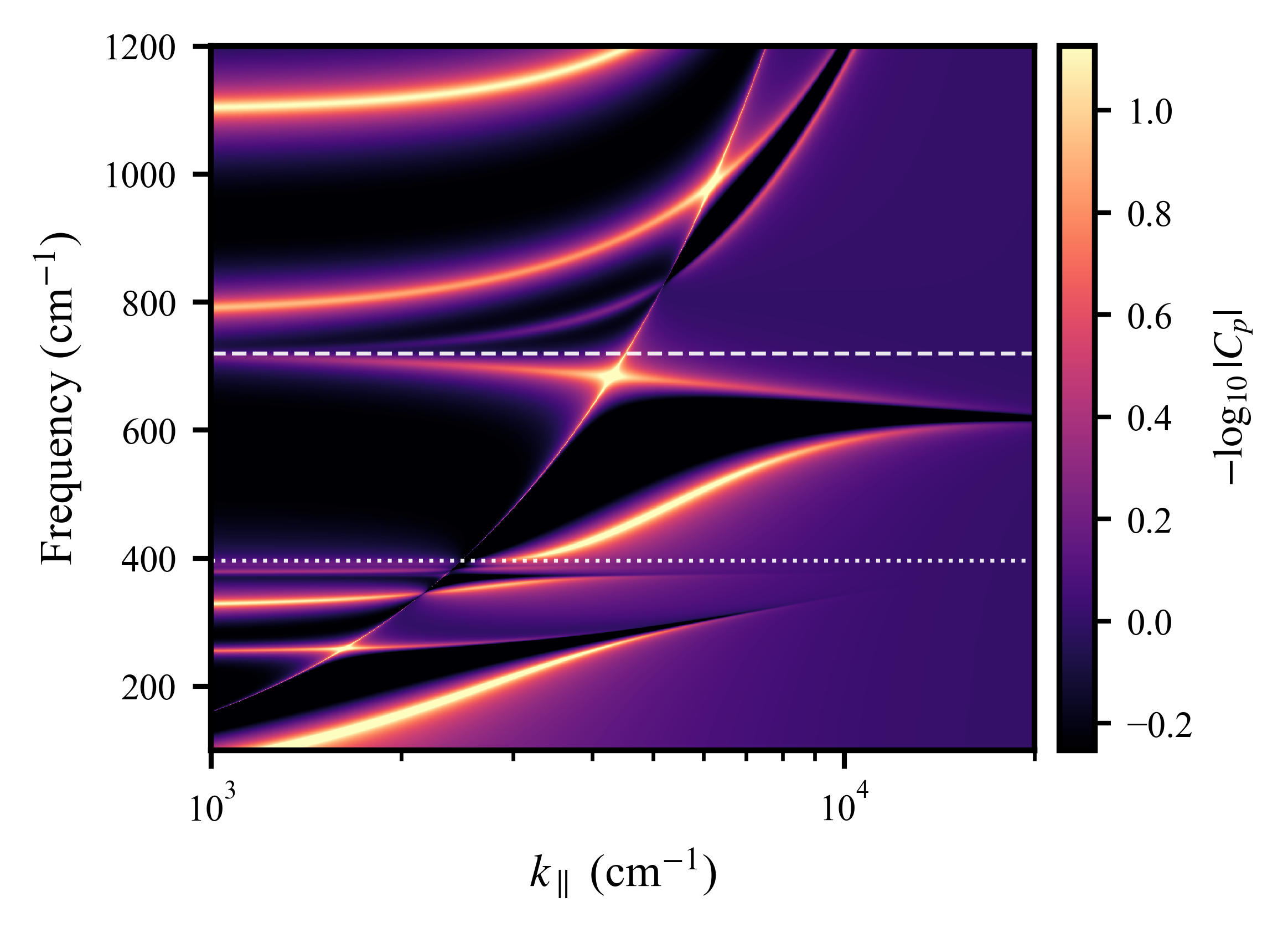}
    \caption{TM polarized dispersion of the Au/MgO multilayer cavity.
    The color scale shows \(\log_{10}|1/C_{\rm TM}|\), where
    \(C_{\rm TM}\) is the TM Fabry-Pérot denominator of the multilayer
    Green tensor (Eq. \ref{eq:fp_denominator}). Bright ridges identify electromagnetic resonances of
    the resonator structure. The horizontal dotted and dashed lines
    mark \(\omega_{\rm TO}\) and \(\omega_{\rm LO}\) of MgO. Below and
    above the Reststrahlen band the dispersive branches are
    predominantly interference or guided modes, whereas the bright branches inside the Reststrahlen band are SPhP modes localized
    near the MgO interfaces.}
    \label{fig:mgo_dispersion}
\end{figure}

Figure~\ref{fig:mgo_dispersion} shows the corresponding TM-polarized
dispersion of the empty Au/MgO multilayer. The resonances are identified
from the zeros of the TM Fabry-Pérot denominator in Eq. \ref{eq:fp_denominator},
\begin{equation}
C_{\rm TM}(\omega,k_\parallel)
=
1-r_{\rm L}^{\rm TM}r_{\rm R}^{\rm TM}
\exp\!\left(2ik_z d_{\rm vac}\right),
\label{eq:tm_denominator}
\end{equation}
so that bright branches in the plot correspond to small
\(|C_{\rm TM}|\). For evanescent waves,
\(k_\parallel>\omega/c\) and \(k_z=i\kappa\), giving
\(\exp(2ik_zd_{\rm vac})=\exp(-2\kappa d_{\rm vac})\). Thus, the
Reststrahlen-band branches are near-field modes whose contribution
decays with distance from the polar interfaces. Outside the Reststrahlen band, the branches are predominantly photon-like interference or guided modes modified by the dispersive multilayer reflection phase. Inside the Reststrahlen band, the branches acquire SPhP character and become strongly evanescent near the MgO interfaces. These near-field modes are
responsible for the large relative DOS,
\begin{equation}
\frac{D_C(\omega)}{D_0(\omega)}
\propto
\frac{
\left\langle
\mathrm{Im}\,\mathrm{Tr}\,
\mathbf G_C(\mathbf r,\mathbf r;\omega)
\right\rangle_{\rm vac}
}{
\mathrm{Im}\,\mathrm{Tr}\,
\mathbf G_0(\omega)
},
\label{eq:relative_dos_green}
\end{equation}
sampled by the molecular absorption spectrum.

To connect these DOS enhancements to molecular kinetics, we compute an absorption-rate-weighted mean photon frequency by averaging molecular absorption frequencies according to the upward radiative rates used in the rigid-rotor anharmonic-oscillator master equation. For a
given microcavity geometry \(L\), we define
\begin{align}
\Omega_{\rm abs}(L,T)
&=
\sum_e k_e^{\rm abs}(L,T),
\\
\langle\omega_{\rm abs}\rangle_L
&=
\frac{1}{\Omega_{\rm abs}(L,T)}
\sum_e \omega_e k_e^{\rm abs}(L,T),
\label{eq:mean_abs_photon_frequency}
\end{align}
where \(e\) runs over all allowed upward absorption events in the coupled oscillator-states. Only upward transitions are included in
\(\langle\omega_{\rm abs}\rangle_L\), since our purpose here is merely to identify the typical photon energy absorbed from the radiation field.

For the \(t_{\rm MgO}=2.5~\mu\mathrm{m}\) series, all values of
\(\langle\omega_{\rm abs}\rangle_L\) fall inside the MgO Reststrahlen
band, between approximately \(495\) and \(522~\mathrm{cm^{-1}}\)
(Table~\ref{tab:mgo_abs_weighted_rate}). 
\begin{table}[t]
\caption{Average absorbed photon frequency and photonic DOS enhancement for the
Au/MgO cavity with \(t_{\rm MgO}=2.5~\mu\mathrm{m}\). The central
vacuum gap is \(d_{\rm vac}=L-2t_{\rm MgO}-2t_s\), with
\(t_s=0.2~\mu\mathrm{m}\). All average absorbed photon frequencies lie
inside the MgO Reststrahlen band, and the decrease of
\(F_{\rm env}(\langle\omega_{\rm abs}\rangle)\) with increasing
\(d_{\rm vac}\) follows the decrease in the relative BIRD rate.}
\label{tab:mgo_abs_weighted_rate}
\begin{ruledtabular}
\begin{tabular}{ccccc}
\(L~(\mu\mathrm{m})\) &
\(d_{\rm vac}~(\mu\mathrm{m})\) &
\(\langle\omega_{\rm abs}\rangle~(\mathrm{cm^{-1}})\) &
\(F_{\rm env}(\langle\omega_{\rm abs}\rangle)\) &
\(k_{\rm cavity}/k_{\rm free}\) \\
\hline
5.5  & 0.1 & 502.58 & 555.82 & 562.36 \\
5.6  & 0.2 & 495.39 & 234.47 & 353.76 \\
5.7  & 0.3 & 495.78 & 138.89 & 273.43 \\
5.8  & 0.4 & 502.18 & 99.81  & 239.61 \\
5.9  & 0.5 & 501.53 & 71.72  & 197.89 \\
6.0  & 0.6 & 498.59 & 53.44  & 162.38 \\
6.5  & 1.1 & 505.54 & 23.99  & 96.05  \\
7.0  & 1.6 & 507.50 & 14.68  & 65.76  \\
7.5  & 2.1 & 511.03 & 10.67  & 50.70  \\
8.0  & 2.6 & 514.57 & 8.49   & 41.59  \\
8.5  & 3.1 & 517.17 & 7.05   & 35.23  \\
9.0  & 3.6 & 519.28 & 6.04   & 30.55  \\
9.5  & 4.1 & 521.04 & 5.31   & 27.05  \\
10.0 & 4.6 & 522.31 & 4.73   & 24.21  \\
\end{tabular}
\end{ruledtabular}
\end{table}
The correlation in Table~\ref{tab:mgo_abs_weighted_rate} shows that the
rate enhancement is controlled by the DOS enhancement at the frequencies actually relevant in radiative absorption. As the central vacuum gap increases from
\(0.1~\mu\mathrm{m}\) to \(4.6~\mu\mathrm{m}\),
\(F_{\rm env}(\langle\omega_{\rm abs}\rangle)\) decreases from
\(555.8\) to \(4.73\), while the relative dissociation rate decreases from \(562\) to \(24.2\). The absorption-weighted frequency changes only weakly and remains inside the Reststrahlen band for the same series of geometries which keep MgO thickness constant. 
Thus, the main geometric effect responsible for reducing enhancement factors with increasing $L$ is not a shift of the dominant molecular
absorption window, but a reduction of the evanescent SPhP DOS used
by that window.

\subsection{Pressure Dependence of Dissociation}

\begin{figure}[t]
    \centering
    \includegraphics[width=\columnwidth]{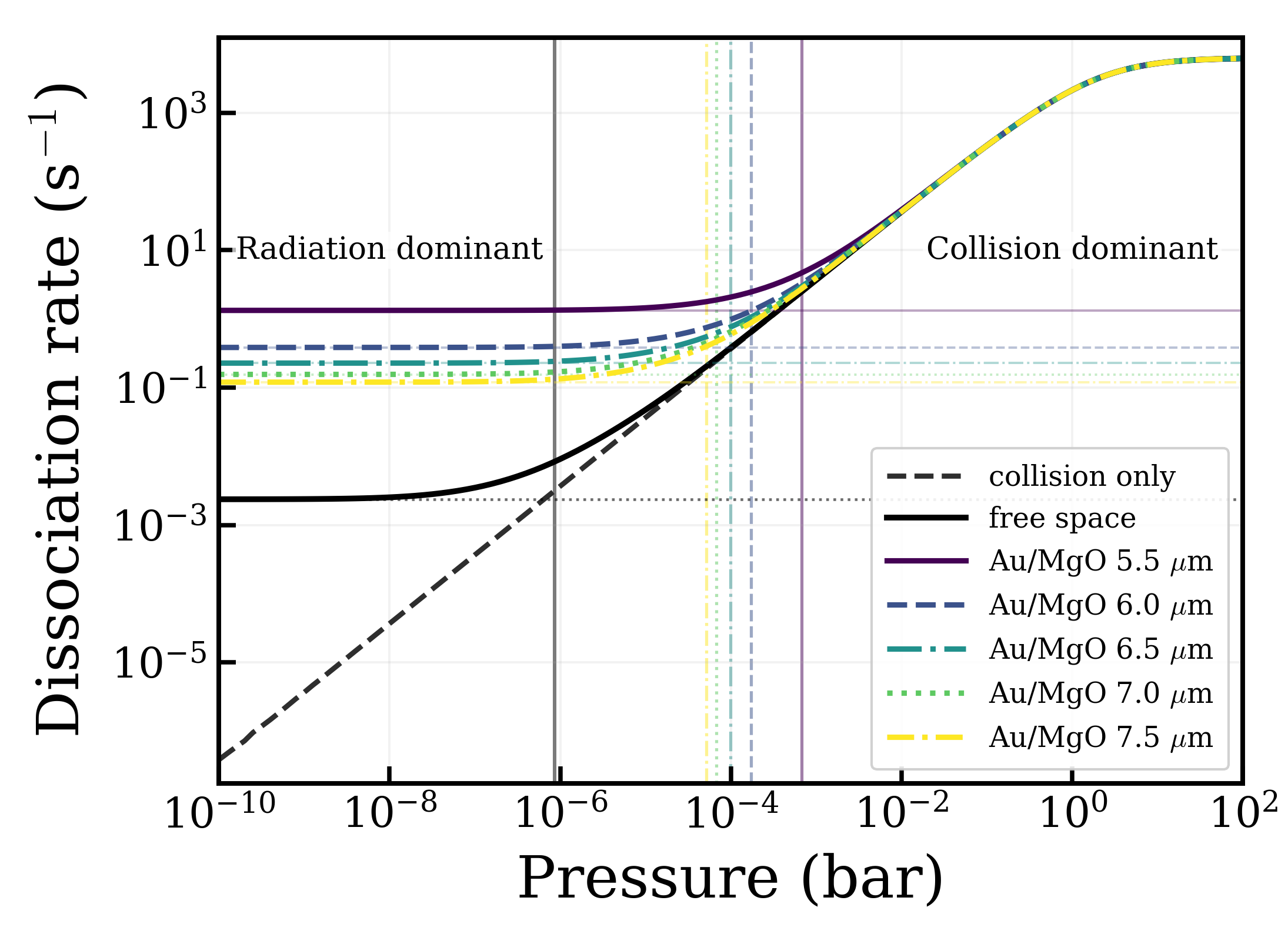}
    \caption{
Dissociation rate \(k_{\rm diss}\) of \ch{(H2O)2Cl-} as a function
of background \ch{CH4} pressure at \(T=300~\mathrm{K}\). The dashed
curve shows the collision only limit, the dotted horizontal lines show
the radiation only baselines, and the solid curves show the full
radiation plus collision master-equation rates for free space and for
Au/MgO cavities with \(L=5.5\), \(6.0\), \(6.5\), \(7.0\), and
\(7.5~\mu\mathrm{m}\). At low pressure, the rates approach
radiation dominated plateaus determined by the free space or
cavity modified photon absorption rates. At high pressure, all curves
approach the collision dominated branch. Vertical dashed lines indicate
the estimated crossover pressures obtained from the
criterion \(P_R=Z/\Omega_{\rm abs}^{\rm cav}\), with the boundary value
\(P_R^\ast\simeq 5\) appropriate for the average absorbed photon energy
in the Au/MgO cavity calculations. The crossover shifts to higher
pressure for resonators with larger \(\Omega_{\rm abs}^{\rm cav}(L,T)\),
showing that microcavity enhanced radiative absorption can significantly extend the radiation dominated regime.
}
    \label{fig:pressure_rrao_au_mgo}
\end{figure}

We next consider the effects of nonzero pressure via the 
pressure-dependent master equation described in
Sec.~\ref{sec:methods_pressure} to determine how background gas
collisions compete with microcavity-modified 
thermal infrared radiative dissociation. Figure
\ref{fig:pressure_rrao_au_mgo} shows the dissociation rate of
\ch{(H2O)2Cl-} at \(T=300~\mathrm{K}\) as a function of the pressure
of the commonly employed \ch{CH4} bath. The dashed gray curve 
is the collision-only result,
the dotted horizontal lines are the radiation-only limits, and the
solid curves are the rates obtained when radiative and collisional
transport are included simultaneously. Free space is shown as a
reference, and the microcavity-modified results correspond to Au/MgO
cavities with \(L=5.5\), \(6.0\), \(6.5\), \(7.0\), and
\(7.5~\mu\mathrm{m}\).

At sufficiently low pressure, the combined rates approach their
radiation-only plateaus. In this limit, collisions are too infrequent
to control the internal-energy flow, so the dissociation rate is set by
radiative activation through the thermal photon field \cite{dunbar1998activation}. The plateau height
therefore depends strongly on the electromagnetic environment. Free
space gives the smallest radiation-only dissociation rate, whereas the Au/MgO
cavities increase the reaction rate by reshaping the photonic density of
states near the molecular transition frequencies as described in the previous subsection. Among the microcavity
lengths shown here, \(L=5.5~\mu\mathrm{m}\) gives the largest
low-pressure rate, while the thicker cavities give progressively weaker
radiative plateaus.

As the pressure is increased, the collision-only branch grows
approximately linearly with pressure. The combined rates then bend away
from their radiation-only plateaus and enter a mixed regime in which
both radiative absorption and collisional energy transfer contribute to
activation. The onset of this mixed regime depends on the electromagnetic environment. A resonator
that produces a larger radiative 
plateau remains radiation dominated to
higher pressure, whereas a weaker 
radiative environment crosses over to
collision-controlled behavior at lower pressure.

At still higher pressure, the combined curves for free space and the
different Au/MgO cavity lengths collapse toward the collision-only
branch. This convergence shows that repeated collisions with the
\ch{CH4} bath eventually dominate the internal-energy redistribution.
In this pressure range, the electromagnetic bath no longer sets
the leading activation rate. Thus, microcavity effects survive only as a secondary correction to the collision-controlled energy distribution.

To quantify the crossover, we adapt the \(P_R\) number criterion
introduced by Zhao et al.  for pressure-dependent unimolecular
kinetics in interstellar environments \cite{Zhao2025Unimolecular}, where infrared radiation and gas-phase collisions can act as competing activation mechanisms. In this case, the relevant dimensionless number is the ratio of the collision frequency \(Z(P,T)\) to the total infrared photon absorption
rate \(\Omega(T)\)
\begin{equation}
    P_R(P,T)=\frac{Z(P,T)}{\Omega(T)}.
\end{equation}
\(P_R\) provides an elementary measure
of whether energy activation is dominated by radiative absorption or by
collisional energy transfer. Radiation-dominated behavior is expected
for \(P_R\lesssim P_R^\ast\), while collision-dominated behavior occurs
for \(P_R\gtrsim P_R^\ast\), where $P_R^\ast$ is the crossover pressure number defined below. In our present scenario, changes in the photonic density of states modify \(\Omega\), and therefore shift the crossover pressure $P_c$ into the collisional dominated regime.

The microcavity modified pressure number is defined as
\begin{equation}
    P_R^{\rm cav}(L;P,T)
    =
    \frac{Z(P,T)}
    {\Omega_{\rm abs}^{\rm cav}(L,T)} ,
    \label{eq:PR_cav_results}
\end{equation}
where \(\Omega_{\rm abs}^{\rm cav}(L,T)\) is the total radiative absorption rate. 

%In a microcavity, the photon absorption rate is no longer a fixed molecular
%quantity because the electromagnetic density of states depends on the
%resonator geometry.

%The crossover pressure is obtained by
%setting the collision to absorption ratio equal to the energy transfer
%boundary appropriate for the present model,
%$
%    Z(P_L^\ast,T)
%    =
%    P_R^\ast\,
%    \Omega_{\rm abs}^{\rm cav}(L,T).
%    \label{eq:PR_cav_boundary_results}
%$
In our Au/MgO computations, the absorption-weighted photon
frequency is approximately
\(\langle\omega_{\rm abs}\rangle\simeq 500~\mathrm{cm^{-1}}\). Using
the average collisional energy transfer estimate
\(\langle\Delta E_{\rm coll}\rangle\simeq 100~\mathrm{cm^{-1}}\), we define
the crossover pressure number as
\begin{equation}
    P_R^\ast
    \simeq
    \frac{\langle\omega_{\rm abs}\rangle}
    {\langle\Delta E_{\rm coll}\rangle}
    \simeq
    \frac{500}{100}
    = 5 .
    \label{eq:PR_star_cavity}
\end{equation}

Hence, using Eq. \ref{eq:collision_frequency},
the microcavity crossover pressure $P_c(L;T)$ can be written as
\begin{equation}
    P_c(L;T)
    =
    \frac{
    P_R^\ast\,
    \Omega_{\rm abs}^{\rm cav}(L,T)\,
    k_{\rm B}T
    }
    {
    x_{\rm bath}k_{\rm coll}(T)
    }.
    \label{eq:cavity_pressure_boundary_results}
\end{equation}
Thus, stronger microcavity enhanced absorption shifts the
radiation dominated regime to higher pressure. This explains the trend
in Fig.~\ref{fig:pressure_rrao_au_mgo}, shorter Au/MgO cavities have
larger radiation only plateaus because they increase
\(\Omega_{\rm abs}^{\rm cav}\), and therefore they also delay the onset
of collision-dominated kinetics.

Overall, Fig.~\ref{fig:pressure_rrao_au_mgo} demonstrates that infrared resonators change not only the zero-pressure BIRD rate but also the
pressure scale at which collisions take over. The microcavity geometry dependence is largest in the radiation-dominated and mixed regimes, where the
radiative rates remain competitive with collisional activation.
In the collision-dominated regime, all environments approach the same
pressure-controlled branch.

\section{Conclusions}

We developed a microcavity-modified master-equation framework for
polyatomic blackbody infrared radiative dissociation in arbitrary electromagnetic environments. The model
combines anharmonic vibrational transition networks, Einstein coefficients, electromagnetic DOS modification, and collisional energy transfer to describe the dissociation kinetics of \(\mathrm{(H_2O)_2Cl^-}\) in structured
infrared resonators. In all computations, the reactive cluster
remains weakly coupled to the electromagnetic field. The microcavity enters by
reshaping the electromagnetic DOS sampled by the molecular transitions.

For the Au/MgO multilayer setup, the largest enhancements originate
from the MgO Reststrahlen response. The polar crystal layers support
surface-phonon-polariton resonances between the transverse and
longitudinal optical phonon frequencies, producing large TM evanescent
contributions to the DOS in the central vacuum reaction region. Across
all Au/MgO geometries considered, the BIRD rate is enhanced for all
microcavity lengths and MgO thicknesses. The enhancement is largest for
short cavities with thick MgO layers, where the molecule samples the
strongest near-field contribution from the polar interfaces. Increasing
the separation between the metallic mirrors weakens this evanescent contribution and monotonically reduces the rate enhancement.

We also examined the competition between radiative
activation and collisional energy transfer with a methane bath gas. At
low pressure, the dissociation rate is controlled by thermal photon
absorption and forms a microcavity-dependent radiative plateau. At higher
pressure, collisional energy transfer dominates and the sensitivity of
the total rate to the resonator DOS is reduced. The environment modified
radiation collision criterion introduced here provides a simple way to
identify radiation-dominated, mixed, and collision dominated regimes and
shows that structured infrared environments can shift the onset of
collision dominated kinetics to significantly higher pressures.

Overall, these results show that infrared microcavities and polaritonic modes emerging from multilayered devices can control polyatomic BIRD by selectively modifying the radiative transport pathways that connect low-energy vibrational states to the dissociation threshold. The strongest effects arise when
phonon-polariton DOS features overlap with kinetically important
anharmonic transitions. This framework provides a practical route for predicting and interpreting microcavity-controlled thermal dissociation of polyatomic ions and suggests that Reststrahlen band materials offer a particularly effective platform for reshaping blackbody driven reaction kinetics.

\begin{acknowledgments}
This work was supported by the donors of the ACS Petroleum Research
Fund under Doctoral New Investigator Grant 66399-DNI6. R.F.R. served as
Principal Investigator on ACS PRF-66399-DNI6, which provided support
for E.S. R.F.R. also acknowledges support from NSF CAREER Award
No.~CHE-2340746 and startup funds provided by Emory University.
\end{acknowledgments}

\bibliography{lib}

\end{document}